\documentclass[aps,prl,preprint,groupedaddress]{revtex4-1}
\usepackage{array}
\usepackage{graphicx}
\usepackage{amssymb}
\usepackage{amsmath}
\usepackage{multirow}
\usepackage{prettyref}
\usepackage{units}
\usepackage[latin1]{inputenc}
\usepackage{amsfonts}
\usepackage{amssymb}
\begin{document}

%\preprint{}

\title{Non-smooth gravity and parity violation}


\author{Iber\^e Kuntz}
\email[]{ibere.kuntz@sussex.ac.uk}
\affiliation{Physics $\&$ Astronomy, University of Sussex, Falmer, Brighton, BN1 9QH, United Kingdom}


%\date{\today}

\begin{abstract}
A conservative extension of general relativity is proposed by alleviating the differentiability of the metric and allowing for non-smooth solutions. We show that these metrics break some symmetries of the Riemann tensor, yielding a new scalar curvature invariant besides the usual Ricci scalar. To first order in the curvature, this adds a new piece of information to the action, containing interesting and unexplored physics. The spectrum of the theory reveals the presence of an additional massless spin-1 field apart from the massless spin-2 graviton. We argue that this new contribution violates P and CP symmetries at leading order in the curvature and we discuss the possibility of observing these effects in existing experiments.
\end{abstract}

%\keywords{}

\maketitle

\textit{Introduction.}---The two pillars of modern physics, quantum field theory and general relativity, are described by fields that are generally assumed to be infinitely differentiable. While this assumption is mathematically appealing, it is not a physical requirement. Here we will show that interesting new physics can be gained by relaxing this unphysical assumption in the gravitational sector.

In addition to the lack of physical motivation for smooth fields, several examples point in the other direction \cite{Baez:2016snx,Dvoeglazov:2002sc}. The desire of symmetrizing Maxwell's equations, for instance, led P. Dirac to conjecture the existence of magnetic monopoles and, consequently, of curves where the electromagnetic potential is singular: the so-called Dirac strings \cite{Dirac:1931kp}. Singularities also play a major role in general relativity through the singularity theorems. Since their discovery, a lot of discussions came up about whether or not they are real and now some people even believe in the existence of such singularities in the universe \cite{Hawking:1973uf}. Furthermore, accommodating shock waves in general relativity requires that the metric be at most Lipschitz continuous \cite{Groah}.

Whether irregularities in the fields are real or just an indication of the failure of our present model remains an open question. In the mean time, while there are no generally accepted quantum theory for gravity that is able to smooth out discontinuities, it is fruitful to try to deal with these imperfections by generalizing general relativity by allowing irregularities from the start (instead of finding them \textit{a posteriori}) and relaxing the differentiability conditions of the theory. This way, one can effectively grasp the physics behind such imperfections even if they are replaced by some smooth structure in a quantum theory.

There has been an increasingly interest in lowering the differentiability of the metric in the context of general relativity. Singularity theorems have been investigated and proven to hold even in the cases of reduced differentiability \cite{Kunzinger:2014fka, Kunzinger:2015gwa}. From a more mathematical point of view, questions concerning differentiability have also been raised regarding well-posedness and motivated by the fact that matter fields need not be smooth and by the possibility of extending local results to global ones \cite{Rendall:2002dr}. There are also instances where Cauchy horizons are not differentiable as commonly believed \cite{Chrusciel:1996tw}.

In this paper, we will relax the smoothness assumption of the metric so that it fails to have continuous partial derivatives of second order, thus failing to belong to the class $C^2$. Note, however, that we are relaxing only the continuity of second derivatives of the metric, but not their existence. We still assume their existence in order to be able to define the curvature of spacetime. Also, note that the continuity of the metric itself and of its partial derivatives of first order remain untouched. Consequently, it follows that the Schwartz theorem does not necessarily hold and partial derivatives of the metric do not commute in general:
\begin{equation}
(\partial_\alpha\partial_\beta-\partial_\beta\partial_\alpha)g_{\mu\nu}(x)=
\begin{cases}
\Delta_{\alpha\beta\mu\nu}(x), & x\in \mathcal{B},\\
0, & x\notin \mathcal B,
\end{cases}
\label{eq:cond}
\end{equation}
where the finite difference $\Delta_{\alpha\beta\mu\nu}(x)$ measures the non-commutativity of second-order derivatives and is related to the size of their jump discontinuity. Note that the non-commutativity in \eqref{eq:cond} only holds at the set of points $\mathcal B$ in spacetime where the derivatives of second order of the metric are not continuous; for regions where the second-order derivatives are continuous (i.e. $x\notin \mathcal B$), they do commute. This type of solution can be obtained, for example, by gluing smooth solutions together. If $g_{\mu\nu}^+$ and $g_{\mu\nu}^-$ are two different solutions of Einstein's equations, then one can construct a new metric $g_{\mu\nu}$ by gluing $g_{\mu\nu}^+$ with $g_{\mu\nu}^-$, which turns out to be non-smooth along the boundary surface $\mathcal B$ between $g_{\mu\nu}^+$ and $g_{\mu\nu}^-$  \cite{zbMATH03127302, Israel:1966rt}. This leads to a novel and natural extension of general relativity that reveals new features of gravity. Here we show that non-smooth metrics allow for the propagation of a spin-1 field that mediates a parity-violating gravitational interaction and which is produced by a varying spin current in the matter sector.

\textit{Non-smooth geometry and curvature invariants.}---Before we begin, it must be stressed that the results from semi-Riemannian geometry might not necessarily hold due to \eqref{eq:cond}, so one must be particularly careful when dealing with this new model. In this section, we will point out the differences produced when such irregularity in the metric is included. The start point remains the same: we are handed with a differentiable manifold $\mathcal M$ equipped with a torsionless and symmetric connection $\Gamma^{\alpha}_{\;\mu\nu}$ and a metric $g_{\mu\nu}$ satisfying \eqref{eq:cond} with a non-zero $\Delta_{\alpha\beta\mu\nu}$. Observe that the manifold continues to be differentiable as we have not changed its differentiable structure, the change only comes in the differentiability of the metric (and of metric-dependent objects) instead. We also assume the metric compatibility of the connection. The definition of the Riemann curvature tensor remains the same. In terms of the metric, it contains terms like
\begin{equation}
R_{\mu\nu\rho\sigma} \supset \frac12 \left(\partial_\rho\partial_\nu g_{\sigma\mu} + \partial_\sigma\partial_\mu g_{\nu\rho} - \partial_\rho\partial_\mu g_{\nu\sigma} - \partial_\sigma\partial_\nu g_{\rho\mu}\right) + \frac12[\partial_\rho,\partial_\sigma]g_{\mu\nu},
\label{eq:rie}
\end{equation}
where $[\partial_\rho,\partial_\sigma] = \partial_\rho\partial_\sigma - \partial_\sigma\partial_\rho$ denotes the commutator of partial derivatives. We can see from \eqref{eq:rie} that the majority of the Riemann tensor symmetries are broken with the exception of the skew-symmetry of the last two indices $R_{\mu\nu\rho\sigma} = -R_{\mu\nu\sigma\rho}$. This implies the existence of different ``Ricci tensors'':
\begin{align}
R_{\rho\sigma}^{(12)} &\equiv g^{\mu\nu}R_{\mu\nu\rho\sigma},\\
R_{\nu\sigma}^{(13)} &\equiv g^{\mu\rho}R_{\mu\nu\rho\sigma},
\end{align}
where the superscript notation is just to remind us the position of the indices of the Riemann tensor that are being contracted, e.g. $R_{\rho\sigma}^{(12)}$ is the result of contracting the first and second indices of the Riemann tensor. All the other contractions of the Riemann tensor are either zero or proportional to the above ones. We still have only one Ricci scalar though, as
\begin{equation}
g^{\rho\sigma}R_{\rho\sigma}^{(12)} = \frac12 g^{\mu\nu}g^{\rho\sigma}[\partial_\rho,\partial_\sigma]g_{\mu\nu} = 0.
\end{equation}
Hence, we define the Ricci scalar to be $R = g^{\nu\sigma}R_{\nu\sigma}^{(13)}$.

With the above differences in mind, we now need to find the curvature invariants. In general relativity, the only curvature invariant that contains up to two derivatives of the metric is the Ricci scalar $R$. However, in the present paper, due to the broken symmetries of the Riemann tensor, another invariant is present. To show this, let us define the dual curvature by \cite{Dvoeglazov}
\begin{equation}
*R_{\mu\nu\rho\sigma} = \frac12 \varepsilon_{\mu\nu}^{\quad\alpha\beta}R_{\alpha\beta\rho\sigma},
\label{eq:dualrie}
\end{equation}
where $\varepsilon_{\mu\nu\rho\sigma} = \sqrt{-g}\epsilon_{\mu\nu\rho\sigma}$ is the Levi-Civita \textit{tensor} and $\epsilon_{\mu\nu\rho\sigma}$ is the Levi-Civita \textit{symbol}. Following \eqref{eq:dualrie}, we define the dual Ricci tensor
\begin{equation}
\widetilde R_{\nu\sigma} = g^{\mu\rho}(*R_{\mu\nu\rho\sigma})
\end{equation}
and the dual Ricci scalar
\begin{equation}
\widetilde R = g^{\nu\sigma}\widetilde R_{\nu\sigma} = \frac12 \varepsilon^{\mu\nu\rho\sigma}R_{\mu\nu\rho\sigma}.
\label{eq:dualrs}
\end{equation}
In general relativity, the dual \eqref{eq:dualrs} vanishes as a consequence of the first Bianchi identity, which does not hold here due to the lack of symmetries of the Riemann tensor. Naturally, the dual Ricci scalar $\widetilde R$ is invariant under diffeomorphisms and, together with the Ricci scalar $R$, they form the basis of invariants that are linear in the curvature.

\textit{Classical non-smooth gravity}.---We aim to build a theory whose symmetry is given by the diffeomorphism group Diff($\mathcal M$); the same symmetry group of general relativity. From the previous section, we are led to formulate the following action
\begin{equation}
S = \frac12\int\mathrm{d}^4x\sqrt{-g}\left(M_p^2 R + M^2 \widetilde R + \mathcal L_M\right),
\label{eq:action}
\end{equation}
where $M_p$ is the Planck mass, $M$ is a mass parameter to be determined by observations and $\mathcal L_M$ represents the matter sector. The action \eqref{eq:action} is the most general action containing up to two derivatives of the metric. The new term in the action above (sometimes named Holst term in the literature \cite{Holst:1995pc}) has been considered previously in the context of torsion gravity \cite{Hojman:1980kv, Shapiro:2014kma}, since some of the symmetries of the Riemann tensor are also broken in the presence of torsion, and as a separate action describing gravitomagnetic physics \cite{Shen:2003pi}. Note, however, that such considerations are drasticaly different from the ones here: while torsion is absent, we allow for non-smooth metrics that contribute with a new term in the gravitational action. In the spirit of effective field theory, one could include higher order Diff($\mathcal M$)-invariant terms in the Lagrangian and then more general dual terms would appear. In this paper, however, we will limit ourselves to first order in the curvature. This ensures the well-defined behavior of the dynamical evolution and opens up the possibility of simulating this theory using the methods from numerical relativity. A more general treatment including higher order dual terms will be the subject of a future work.

From the variation of Equation \eqref{eq:action}, one obtains the equations of motion
\begin{equation}
R_{\mu\nu}-\frac12 g_{\mu\nu}R + \frac{M^2}{M_p^2}(\widetilde R_{\mu\nu} - \widetilde R_{\nu\mu}) = \frac{1}{M_p^2}\Theta_{\mu\nu},
\label{eq:eom}
\end{equation}
where $\Theta_{\mu\nu}$ is the canonical and non-symmetric energy-momentum tensor. For simplicity, we are using the notation $R_{\mu\nu}\equiv R_{\mu\nu}^{(13)}$ as $R_{\mu\nu}^{(13)}$ is the only Ricci tensor that will appear in the rest of this article. The non-symmetric tensor $\Theta_{\mu\nu}$ can be split into its symmetric and anti-symmetric part by reversing the Belinfante-Rosenfeld procedure \cite{BELINFANTE1940449}
\begin{equation}
\Theta_{\mu\nu} = T_{\mu\nu} - \frac12 \nabla_\alpha(S_{\nu\mu}^{\ \ \alpha}+S_{\mu\nu}^{\ \ \alpha}-S^\alpha_{\ \nu\mu}),
\label{eq:em}
\end{equation}
where $S_{\mu\nu}^{\ \ \alpha}$ denotes the spin current. Note that the parameter $M$ cannot be calculated by comparing the non-relativistic limit of the theory with the Newtonian potential as this procedure amounts on taking the component $R_{00}$ of the Ricci tensor, killing the anti-symmetric part of the equation of motion altogether. Any smooth metric would also yield a vanishing dual Ricci tensor, turning all general relativity solutions into solutions of \eqref{eq:eom} for any value of $M$.

Equation \eqref{eq:eom} can be conveniently separated into its symmetric and anti-symmetric part
\begin{align}
\label{eq:eoms}
&R_{(\mu\nu)} - \frac12 g_{\mu\nu}R = \frac{1}{M_p^2}T_{\mu\nu},\\
\label{eq:eoma}
&R_{[\mu\nu]} + \frac{M^2}{M_p^2}(\widetilde R_{\mu\nu} - \widetilde R_{\nu\mu}) = \frac{1}{M_p^2}\nabla_\alpha S^\alpha_{\ \mu\nu},
\end{align}
where $(\cdot)$ and $[\cdot]$ denote index symmetrization and anti-symmetrization, respectively. Equation \eqref{eq:eoms} looks like Einstein's equations with the Ricci tensor replaced by the symmetrized Ricci tensor and, in fact, it recovers general relativity for smooth metrics. Its source is given by the usual Einstein-Hilbert energy-momentum tensor. On the other hand, the source for the anti-symmetric Equation \eqref{eq:eoma} is due to spin currents. This could bring important consequences to quantum gravity as it shows that spins also gravitate. Equation \eqref{eq:eoma} is entirely due to the non-smooth solutions and, for this reason, we will refer to it by non-smooth sector.

All that has been done so far concerning classical general relativity involves Equation \eqref{eq:eoms} only. Our belief in the smoothness has made us ignore Equation \eqref{eq:eoma} completely, leaving the non-smooth sector unexplored. It is interesting to observe that the extension to non-smooth metrics composes a modification of general relativity in the low energy limit without messing with the success of the Einstein-Hilbert term in explaining most of the gravitational phenomena. In fact, $\widetilde R$ is non-vanishing only for non-smooth metrics and for smooth ones Equation \eqref{eq:action} reduces to the Einstein-Hilbert action. Nonetheless, the non-smooth sector opens a new door for the exploration of gravitational phenomena that have been hiding in the non-smooth structures all along. Not only this could generate answers to the present problems, but also could lead to the discovery of new effects. In the next sections, we will describe some of these new effects.

\textit{Linearized theory and gravitational waves}.---The first question that arrises is whether non-smooth metrics give rise to new degrees of freedom that after quantization would lead to new particles. To address this point, we will linearize the vacuum field equations around Minkowski spacetime $g_{\mu\nu} = \eta_{\mu\nu} + h_{\mu\nu}$. In the absence of matter, Equation \eqref{eq:eoms} becomes
\begin{equation}
\label{eq:eomsv}
R_{\mu\nu} = -R_{\nu\mu}
\end{equation}
and thus we find that the Ricci tensor is completely anti-symmetric. Linearizing Equation \eqref{eq:eomsv} gives
\begin{equation}
\partial^\mu\partial_\sigma\bar h_{\mu\nu} + \partial^\mu\partial_\nu \bar h_{\mu\sigma} - \Box\bar h_{\nu\sigma} + \frac12\eta_{\nu\sigma}\Box\bar h = 0,
\end{equation}
where $\bar h_{\mu\nu} = h_{\mu\nu}-\tfrac12 h\eta_{\mu\nu}$. The usual Hilbert gauge $\partial^\mu \bar h_{\mu\nu}=0$ is of no help to us here because of the non-commuting derivatives. However, one can still find a gauge where
\begin{equation}
\label{eq:ghilbert}
\partial^\mu\partial_\sigma\bar h_{\mu\nu}=0,
\end{equation}
which will be called generalized Hilbert gauge. In fact, under the infinitesimal coordinate transformation $x^\mu\to x'^\mu = x^\mu + \xi^\mu(x)$, we have
\begin{equation}
\partial^\mu\partial_\sigma\bar h_{\mu\nu} \to (\partial^\mu\partial_\sigma\bar h_{\mu\nu})' = \partial^\mu\partial_\sigma\bar h_{\mu\nu} - \Box\partial_\sigma\xi_\nu.
\end{equation}
Thus any solution of $\partial^\mu\partial_\sigma\bar h_{\mu\nu} = \Box\partial_\sigma\xi_\nu$ for $\xi^\mu$ gives the desired gauge in the new coordinates. Analogously, one can also prove that, outside the source, we can set
\begin{equation}
\label{eq:tt}
\bar h = 0,\quad \bar h^{0\mu} = 0,
\end{equation}
without spoiling the gauge $\partial^\mu\partial_\sigma\bar h_{\mu\nu}=0$. This is analagous to the transverse traceless (TT) gauge, where we have a generalized form of the Hilbert gauge adapted to our needs. Hence, using these choices of gauge we get
\begin{equation}
\Box h_{\mu\nu} = 0.
\label{eq:wave}
\end{equation}
We should stress that \eqref{eq:wave} is not simply the old and well-known gravitational wave equation of general relativity as the field $h_{\mu\nu}$ is now allowed to be non-smooth. This proves that both smooth and non-smooth metric solutions are waves propagating in spacetime.

Now let us see what the anti-symmetric field equation \eqref{eq:eoma} has to say at the linear level. By using $\bar h = 0$, we obtain
\begin{equation}
\epsilon_\mu^{\ \sigma\alpha\beta}\partial_\sigma\partial_\beta h_{\alpha\nu} - \epsilon_\nu^{\ \sigma\alpha\beta}\partial_\sigma\partial_\beta h_{\alpha\mu} = 0.
\label{eq:linear2}
\end{equation}
Attention to the fact that in the linear equations only the Levi-Civita symbols appear as the Levi-Civita tensor only contributes to higher order terms in $h$. Defining a tensor $A_{\mu\nu}$ by
\begin{equation}
\partial^\sigma A_{\mu\nu} = \epsilon_\mu^{\ \sigma\alpha\beta}\partial_\beta h_{\alpha\nu},
\end{equation}
allows us to write \eqref{eq:linear2} as
\begin{equation}
\Box B_{\mu\nu} = 0,
\end{equation}
where $B_{\mu\nu} = A_{\mu\nu} - A_{\nu\mu}$. Therefore, the anti-symmetric equations of motion lead to the existence of an additional massless anti-symmetric rank-2 field. Although it is tempting to claim that this field has spin 2, it is actually a spin-1 field. In fact, we remind the reader that every anti-symmetric field can be decomposed as
\begin{equation}
\label{eq:reps}
B_{\mu\nu}\in \mathbf{1} \oplus \mathbf{1}
\end{equation}
showing it contains two spin-1 representations. However, there is an apparently mismatch as massless representations of the Poincar\'e group has only two degrees of freedom, while in principle $B_{\mu\nu}$ has six. This difference is due to spurious degrees of freedom that are removed after fixing a gauge. In fact, by using the gauge choices \eqref{eq:ghilbert} and \eqref{eq:tt}, we obtain $\partial^\nu\partial^\sigma B_{0\nu} = 0$, which yields four constraints, reducing the number of degrees of freedom to two, as expected for a massless field. Therefore, by fixing the gauge we removed one of the spin-1 representations of \eqref{eq:reps} entirely and we conclude that $B_{\mu\nu}$ describes a massless spin-1 field. The spectrum of theory is thus composed by two massless fields: one of spin 1 and another of spin 2. Note that the spin-1 field $B_{\mu\nu}$ lives only in the submanifold $\mathcal B\subset \mathcal M$ of the spacetime where the second-order derivatives of the metric have jump discontinuities, vanishing identically outside $\mathcal B$. On the other hand, the spin-2 field $h_{\mu\nu}$ exists everywhere in $\mathcal M$.

\textit{Field's production}.---To discuss the production of the field $B_{\mu\nu}$ we just need to put the matter back in. The solution is
\begin{equation}
B_{\mu\nu}^{TT} = \frac1{M^2}\int\mathrm{d}^4x' G(x-x') \partial_\alpha S^\alpha_{\mu\nu}(x'),
\label{eq:sol}
\end{equation}
where $G(x-x')$ is the retarded Green's function of the wave operator and $B_{\mu\nu}^{TT}$ is the field's projection to the TT modes. If the distance from the detector to the source $r=|x-x'|$ is much larger than the source radius, to leading order Equation \eqref{eq:sol} becomes
\begin{equation}
B_{\mu\nu}^{TT} = \frac{1}{8\pi M^2 r}\frac{d\Sigma_{\mu\nu}}{dt}\Bigr|_{t_\text{ret}},
\end{equation}
where $\Sigma_{\mu\nu}$ is the spin angular momentum and $t_\text{ret}$ is the retarded time. We see that the field $B_{\mu\nu}$ is produced by a time variation of the spin angular momentum. For most types of macroscopical matter, the constituent spins point at arbitrary directions, thus the total contribution to the spin angular momentum at macroscopic scales (i.e. outside matter) averages out. Of course this is only true when there is no effect capable of lining up the spins. Neutron stars, for example, are potential physical systems for the exploration of the spin-1 waves since they possess strong magnetic fields able to produce a non-zero net spin angular momentum. However, it is important to stress that $B_{\mu\nu}$ is produced by the variation of the spin angular momentum, thus we must seek a rapid-varying magnetic field and not only a strong one.

A varying magnetic field can be easily produced in a controlled manner by using ferromagnetic materials here on earth. A setup similar to the Einstein-de Haas experiment \cite{einsteindehaas}, where one is able to control the variation of the spin current of a suspended ferromagnetic object, should be sufficient to at least constrain the parameter $M$. For example, one could use two adjacent copies of the Einstein-de Haas setup, i.e. two pendulums consisting of ferromagnetic objects suspended by a string that are brought together side-by-side. Each ferromagnetic object is then embedded into different external magnetic fields whose magnitude and direction can be controlled separatelly. Varying the magnetic fields ultimately leads to variations in the spin current of the ferromagnetic materials as the spins are forced to line up with the external magnetic fields. The variation of spin currents would then create a relative force, intermediated by $B_{\mu\nu}$, between the suspended ferromagnetic objects, which would make these objects move along the line that crosses their centers (a force between spin currents have also been studied in \cite{Naik:1981ka,Pradhan:1985jh}). One could then measure the angular variation of the pendulum (as opposed to the angle of twist of the original Einstein-de Haas experiment) with respect to the vertical axis, which consequently gives a measurement of the spin-1 force. Note that, at the present moment, there is no data for the force just described as the original Einstein-de Haas experiment did not aim to measure this interaction.

\textit{Parity violation}.---Since the experimental discovery of parity violation in the weak sector, discrete symmetry breaking has been the subject of several studies in the particle physics community, particularly with the hope of finding symmetry violations in the other sectors. In QCD, for example, it was found natural terms in the Lagrangian that could break CP. However, such violation has never been observed, which led to the infamous Strong CP problem. The Chern-Simons modified gravity also suggests the existence of CP-violating interactions, this time in the gravitational sector \cite{Jackiw:2003pm,Alexander:2009tp}.

The second term in the action \eqref{eq:action} features a similar property. According to \eqref{eq:dualrs}, $\widetilde R$ is obtained by contracting the Levi-Civita tensor with the Riemann curvature. The Levi-Civita tensor, in turn, transforms as a tensor under diffeomorphisms connected to the identity, but under transformations with negative Jacobian it changes its sign as $\varepsilon^{\mu\nu\rho\sigma}\to\varepsilon'^{\mu\nu\rho\sigma}=-\varepsilon^{\mu\nu\rho\sigma}$. The reason is that the Levi-Civita tensor depends on the orientation chosen, thus orientation-reversed transformations (i.e. transformations with negative Jacobian) flip its sign. As a corollary, the term $\widetilde R$ in the Lagrangian leads to parity (P) and time-reversal (T) symmetry breaking. From the CPT theorem, T violation implies CP violation so that CPT remains unbroken. Therefore, non-smooth gravitational solutions are able to produce P and CP violations in the gravitational sector, a feature that is not present in general relativity. Note that the standard model fields couple to the CP-violating field $B_{\mu\nu}$ via the divergence of their spin current, i.e. $\partial_\alpha S^\alpha_{\ \ \mu\nu}$. Therefore, the strength of $B_{\mu\nu}$ (and thus the degree of CP violation) ultimately depends on the free parameter $M$ and on $\partial_\alpha S^\alpha_{\ \ \mu\nu}$. The smaller $M$ or the higher $\partial_\alpha S^\alpha_{\ \ \mu\nu}$, the bigger will this effect be.

While the idea of having parity violation in the gravitational sector is not new, the main difference between our finding and Chern-Simons gravity is that in our case P and CP are violated at leading order in the curvature, while the Chern-Simons modification shows up only at higher order and is expected to be Planck supressed in low energies. This brings the possibility of observing more easily such violations in present or near-future experiments as they now appear at much more accessible energies.

Independently of the energy scale where parity is violated, various tests of gravitational P violation have been proposed in the past few years. In particular, P violation in the gravitational sector generates an asymmetry in the polarization of gravitational waves, known as gravitational wave amplitude birefringence, that gets amplified after the propagation over cosmological distances \cite{Lue:1998mq,Yunes:2010yf,Contaldi:2008yz}. The amplified signal could then be measured in the CMB \cite{Balaji:2003sw}, for example. Supermassive binary black holes have also been suggested as a potential physical system to probe gravitational parity violation with LISA \cite{Alexander:2007kv}. In \cite{Yunes:2010yf} was argued that even existing instruments, such as LIGO and VIRGO, could be able to test gravitational parity violation through the detection of coincident gravitational waves events. Therefore, given the fact that P violation should be present at very low energies according to our theory, we believe that gravitational parity violation will soon be found.

\textit{Conclusions}.---We showed in this paper how a simple and natural extension of general relativity, where the smoothness requirement of the metric is abandoned, opens up the possibility for interesting phenomena. The theory agrees with all observations up to date as it has general relativity as the subset in which the metric is smooth. In particular, all the known solutions of general relativity, including Schwarzchild, Kerr, FLRW, de Sitter, anti-de Sitter and so on, are still exact solutions in this particular modification. The differences only come up when considering non-smooth solutions, which has not been probed experimentally yet as all the observations were aiming to the particular set of known solutions. By opening this black box, we showed that an anti-symmetric massless spin-1 field comes into play along with the standard massless spin-2 graviton. We also argued that P and CP symmetries are violated due to the new term in the Lagrangian. This violation is mediated by the anti-symmetric field and it takes place at the same scale as the standard general relativistic effects, as opposite to the Planck supressed Chern-Simons gravity violation. This makes its observation a lot easier and, together with the violations from other sectors, it could result in a possible explanation for the matter-antimatter asymmetry through baryogenesis. Such violations might be observed in the CMB or in experiments such as VIRGO, LIGO and LISA, through the generation of a birefringence in the gravitational waves. It is important to stress, however, that none of the fundamental symmetries (diffeomorphism and local Lorentz invariance) were broken. Neither are drastical changes being made. We are only allowing for not so well-behaved metrics, which according to the authors point of view are still a very conservative extension of general relativity and, in fact, even more conservative than assuming a smooth metric without any good reason.

\begin{acknowledgments}
This work is supported by the National Council for Scientific and Technological Development (CNPq - Brazil).
\end{acknowledgments}

\end{document}